\documentclass[aps,pra,twocolumn,showpacs,superscriptaddress,groupedaddress]{revtex4} 
\usepackage{graphicx} 
\usepackage{dcolumn} 
\usepackage{bm} 
\usepackage{amssymb} 
\usepackage{braket} 

\begin{document}
\widetext

\title{High teleportation rates using cold-atom ensemble based quantum repeaters with Rydberg blockade}
\author{Neal Solmeyer}
\affiliation{Quantum Sciences Group, Army Research Laboratory, 2800 Powder Mill Rd., Adelphi, Maryland 20783, USA}
\author{Xiao Li}
\affiliation{Joint Quantum Institute and Department of Physics, University of Maryland, College Park, Maryland 20742, USA}
\author{Qudsia Quraishi}
\affiliation{Quantum Sciences Group, Army Research Laboratory, 2800 Powder Mill Rd., Adelphi, Maryland 20783, USA}

\date{\today}

\begin{abstract}
We present a simplified version of a repeater protocol in a cold neutral-atom ensemble with Rydberg excitations optimized for two-node entanglement generation and describe a protocol for quantum teleportation. Our proposal draws from previous proposals [Zhao, {\sl et al.}, Phys. Rev. A {\bf 81}, 052329 (2010)] and [Han, {\sl et al.} Phys. Rev. A {\bf 81}, 052311 (2010)] who described efficient and robust protocols for long-distance entanglement with many nodes. Using realistic experimental values we predict an entanglement generation rate of $\sim25$ Hz and teleportation rate of $\sim$ 5 Hz. Our predicted rates match the current state of the art experiments for entanglement generation and teleportation between quantum memories. With improved efficiencies we predict entanglement generation and teleportation rates of $\sim$7.8 kHz and $\sim$3.6 kHz respectively, representing a two order of magnitude improvement over the currently realized values. 
Cold-atom ensembles with Rydberg excitations are promising candidates for repeater nodes because collective effects in the ensemble can be used to deterministically generate a long-lived ground state memory which may be efficiently mapped onto a directionally emitted single photon. 
\end{abstract}

\maketitle
\section*{\label{sec:level1}INTRODUCTION}

Quantum repeaters can be used to entangle remote quantum memories by interfering flying qubits that are in turn entangled with the memories\cite{Briegel1998}. Networks of quantum repeaters are capable of extending the distance of quantum communication beyond what is capable in purely photonic systems by dividing the communication channel into smaller segments with a quantum memory at each node\cite{Sangouard2011}. In addition, a network of quantum memories could be enabled with a quantum register of multiple memories at each node for logical operations and error correction\cite{VanMeter2014}. This type of quantum network could realize applications such as cluster state generation\cite{Zwierz2009}, distributed quantum computation\cite{Eisert2000}, and entanglement enhanced measurement\cite{Komar2014, Gottesman2012}. In order for the entanglement to be distributed to remote sites, it is desirable for the nodes of a network to include long-lived quantum memories entangled with photonic flying qubits for long-distance communication. Quantum teleportation\cite{Bennett1993} is a vital protocol to realize on such quantum networks because it allows the transmission of an unknown quantum state from one node to another while still adhering to the no-cloning theorem \cite{Wooters1982}. 

In this paper we present a protocol for teleportation between quantum repeater nodes based on Rydberg excitations in neutral atom ensembles. We describe the protocol in detail and examine the performance of entanglement generation and teleportation protocols for two-nodes. The photon collection is enhanced by using collective effects in the ensemble for directional photon emission. The fidelity of each step in a many step protocol can limit the success rate of the protocol. The use of Rydberg blockade allows us to improve the fidelity of each step, particularly the memory generation step, over processes that rely solely on spontaneous emission. The entanglement generation protocol we use is a modified version of that proposed in the previous work of Zhao {\sl et al.}\cite{Zhao2010}. The optimization for our two-node protocol minimizes the number of steps and ground states needed which in principle improves the probability of success, reduces the time needed for the protocol and improves the state fidelities. Our two-node optimization comes at the expense of the many-node scaling characteristics of the previously proposed protocols. We compare our protocol with those of \cite{Zhao2010} and \cite{Han2010} and find that a simplified protocol can be advantageous for two node protocols because it has less experimental overehead at each node and higher rates for small rapid networking of small numbers of nodes. 

Using experimental values that have been achieved in similar systems, we predict that entanglement generation and teleportation rates of $\sim$25 Hz and $\sim$5 Hz are possible. If technology such as pulse shaping, three dimensional optical lattices, or the use of optical cavities is used, it is reasonable to predict that one could achieve experimental efficiencies that lead to entanglement generation rates as high as $\sim$7.8 kHz with corresponding teleportation rates $\sim$3.6 kHz. This would represent asignificant improvement over the highest currently achieved rates for two-node protocols with memory\cite{Nolleke2013, Hucul2015}. 

Teleportation between matter nodes has been realized in ions\cite{Olmschenk2009}, neutral atoms\cite{Bao2012} and most recently in NV centers\cite{Pfaff2014}. These examples rely on spontaneous emission from an excited state to generate the memory state. Because of a combination of probabilistic memory generation and low photon collection efficiency, teleportation between quantum memories has generally had low rates, on the order of one every few minutes. Approaches to achieving Hz-level rates have included custom high numerical aperture collection lenses\cite{Hucul2015} or placing the quantum memory in an optical cavities \cite{Nolleke2013}. 

Photonic systems can also be used for quantum teleportation\cite{Bouwmeester1997, Ma2012}, and secure quantum key distribution protocols\cite{Scarani2009}. Photonic systems could realize applications in quantum communication and computation with cluster states through one-way measurement-based computation\cite{Walther2005, Chen2014}. However, purely photonic systems may be limited because of the difficulties associated with the distance limitation from exponential photon loss in an optical fiber and the incorporation of information processing of multiple qubits at each node. A quantum network enabled with quantum memories addresses both of these difficulties.

The paper is structured as follows. First, neutral atom based quantum repeaters are introduced in Section~\ref{sec:level1.1}. The process for producing ground state memories entangled with directionally emitted photons is detailed in Section~\ref{sec:level2}. In Section~\ref{sec:level3.0} we discuss the proposed system for experimental realization. In Section~\ref{sec:level3} we show a simplified version of the protocol for generating entangled flying qubits based on the protocols in Zhao {\sl et al}. \cite{Zhao2010}. In Section~\ref{sec:level4} we show how the entanglement can be generated between remote memory pairs. In Section~\ref{sec:level5} we demonstrate a theoretical protocol for quantum teleportation within this framework. In Section ~\ref{sec:level5.5} we analyze the entanglement generation and teleportation rates for two node protocols in this system. Finally, in Section~\ref{sec:level6} we analyze a model for the many node entanglement generation and teleportation rates.

\section{\label{sec:level1.1}Neutral atom based quantum repeaters}

The Duan, Lukin, Cirac, and Zoller (DLCZ) proposal \cite{Duan2001}theoretically described a realizable repeater protocol based on directional single-photon emission from a neutral atom ensemble that relies only on linear optics. The DLCZ protocol uses weak laser beams to probabilistically excite a single spin-wave in an ensemble. The spin-wave serves as the quantum memory and can be read out with a subsequent strong pulse. A phase matching condition, similar to that in four-wave mixing, provides a collective enhancement in the photon emission direction and ensures that the single photons can be efficiently captured. However, in order to reduce two-photon errors, the probability of exciting a single spin wave quantum memory and generating the heralding photon, or the `write' photon, must be kept to $\sim 10^{-3}$ or lower, leading to low rates of entanglement generation\cite{Kuzmich2003, vanderWal2003, Sangouard2011}

In contrast, for the case of neutral atoms, the Rydberg blockade mechanism offers a route to improve these rates by generating the quantum memory deterministically\cite{Lukin2001, Saffman2002}. Rydberg excitations in ensembles can utilize collective enhancement from phase matching, similar to the DLCZ scheme, to ensure efficient collection of a single photon entangled with a quantum memory. This can in principle increase the rate of successfully generating a quantum memory as compared to the DLCZ protocol by three orders of magnitude.

Entanglement between a collective Rydberg excitation and a single photon was demonstrated experimentally in Li {\sl et al.}\cite{Li2013}. The single photon was entangled with a quantum memory by using a partial readout of the Rydberg level. However, using the Rydberg level as the memory limits the lifetime to a few tens of $\mu$s. 

To improve the memory lifetime, the Rydberg excitation can be shelved in a long-lived atomic ground state. Shelving single collective excitations into ground states via Rydberg states was experimentally demonstrated by \cite{Ebert2014} with lifetimes as long as a few ms \cite{Ebert2015}. To increase the coherence time of Zeeman state memories, it may be possible to adapt methods that have been used to increase the coherence times of ground state memories to several seconds\cite{Dudin2013}. Extraction of a single photon entangled with a long-lived ground-state quantum memory via the Rydberg blockade mechanism has not yet been realized experimentally. 

The Rydberg blockade mechanism can also be used to perform two-qubit gates, which are a critical component of the entangling and teleportation protocols described below. A two-qubit gate using Rydberg blockade between two neutral atoms has been demonstrated experimentally\cite{Isenhower2010}. In addition, Rydberg blockade between two ensembles has been demonstrated\cite{Ebert2015}. The ability to efficiently perform deterministic gates between local qubits can allow for advanced protocols such as error correction. 

Zhao {\sl et al.}\cite{Zhao2010} and Han {\sl et al.}\cite{Han2010} developed protocols for quantum repeaters using cold atom ensembles with Rydberg excitations that have favorable scaling to long distances and many nodes. In these protocols, the memories are deterministically generated via Rydberg blockade. Multiple memories are stored by coherently driving single excitations to different Zeeman ground states. Gates are performed between the multiple memories using Rydberg blockade. Entanglement is generated between remote pairs of nodes through photon interference. In addition to the memory generation, Rydberg blockade is used for deterministic entanglement swapping through local Rydberg interactions rather than photon interference and detection, which is typically probabilistic and of low efficiency. A separate proposal includes coupling the atomic ensembles to optical cavities to make use of the high efficiency photon absorption in a cavity in order to generate remote entanglement without the need for photon detection \cite{Brion2012}. 

Because of their potentially high rates of communication and information processing capabilities, quantum repeater nodes based on neutral atom ensembles with Rydberg excitations have the potential to enable large-scale quantum networks. Our work aims to simplify these protocols for small numbers of nodes and flesh out the details in order to pave the way for initial experimental demonstrations.

\section{\label{sec:level2}Single photon storage and readout}

\begin{figure}
\includegraphics[width=1.0\columnwidth]{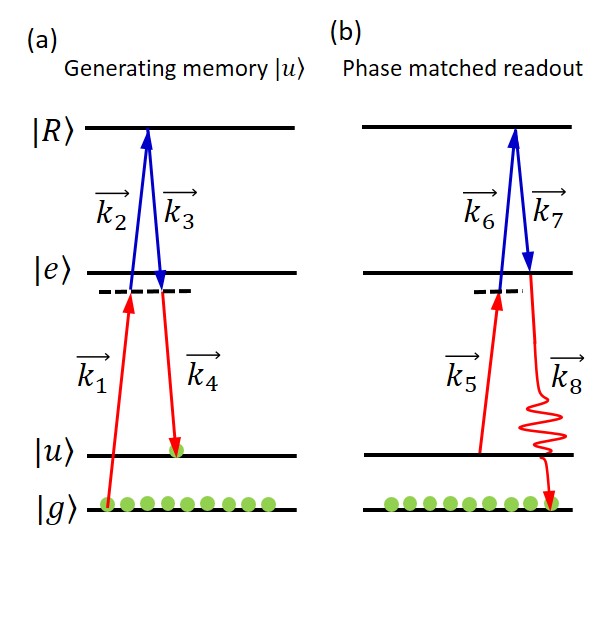}
\caption{\label{fig:figone}The level scheme shows the reservoir state $\Ket{g}$ on the same hyperfine manifold as the memory state $\Ket{\bf u}$. Both $\Ket{g}$ and $\Ket{\bf u}$ are coupled with two-photon transitions to a Rydberg state $\Ket{\bf R}$ through an intermediate excited state $\Ket{\bf e}$. (a) shows the generation of a single memory state, and (b) shows the readout from the Rydberg level. Collective enhancement occurs when the spontaneously emitted photon $\vec{\bf{k_8}}$ brings the state back to the reservoir $\Ket{g}$.}
\end{figure}

Here we examine the process of writing a single quantum memory and efficiently mapping it onto a photon mode. Consider a simplified atomic structure as in Fig.~\ref{fig:figone} (a) with two $^{87}$Rb ground states $\Ket{g}$ and $\Ket{u}$, an excited state $\Ket{e}$ which, for example, could be in the $P_{3/2}, F=2$ manifold, and a high lying Rydberg level $\Ket{R}$. The atoms are initially optically pumped into the $\Ket{g}$ state which serves as the reservoir state where most atoms will remain. We couple the ground states to $\Ket{R}$ with a two-photon processes through the intermediate excited state $\Ket{e}$, as shown in the the energy-level diagram in Fig.~\ref{fig:figone}. For a two-photon transition, detuning from the intermediate state ensures minimal population is transferred to $\Ket{e}$ during the process and two-photon resonance ensures that the state is transferred with high fidelity. 

Due to the large interactions between Rydberg states, the presence of one Rydberg excitation will shift the Rydberg energy levels of near-by atoms out of resonance with the excitation beams. This ensures that only one excitation occurs within a Rydberg radius, $r_b$. This radius depends on the target Rydberg states and the linewidth of the excitation laser. If the trapped atomic ensemble has a diameter smaller than $r_b$ and the excitation beams are significantly larger than the atomic ensemble, then the excitation beams interact with all atoms with equal strength and the ensemble can contain only one Rydberg excitation.

When a two-photon $\pi$-pulse is applied, i.e. $\vec{\bf{k_1}}$ and $\vec{\bf{ k_2}}$ in Fig.~\ref{fig:figone} (a) we assume that each atom has an equal probability of being excited, as is discussed in Section~\ref{sec:level3.0}. The resulting state is an equal superposition of one atom in the excited Rydberg state with the remaining atoms in the reservoir state $\Ket{g}$. The collective state of the ensemble is a $\Ket{W}$ state and has the form given in Eq.~(\ref{eq:super}). 

\begin{equation}
\Ket{\bf{R}}=\frac{1}{\sqrt{N}}\sum_{j=1}^N e^{i\phi_j}\Ket{g...R_j...g} 
\label{eq:super}
\end{equation}
Where $ N $ is the total number of atoms within $r_b$. The phase, $\phi_j$, is determined by the wave vectors of the excitation laser beam and the position of the atom, i.e. ${ \phi_j =\sum^{pulses}\vec{\bf{k}}\cdot \vec{\bf{r_j}}}$. 

Because the excitation is shared across the atoms, the state is robust against atom loss and the effective Rabi frequency of the two-photon transition between $\Ket{g}$ and $\Ket{R}$ is enhanced by a factor of $\sqrt{N}$ compared to the single-atom Rabi frequency\cite{Lukin2001, Saffman2002}. 

Importantly, when generated with the Rydberg blockade mechanism there is no vacuum component in the produced $\Ket{W}$ state, and therefore it can be prepared deterministically. In addition, the two-photon component depends on the detuning from the Rydberg blockade shifted state, which can be made to lead to low two-photon errors\cite{Sangouard2011}.

This is in contrast to a $\Ket{W}$ state generated by the DLCZ protocol which produces a state that is mostly in the vacuum state, that is, almost all of the atoms remain in $\Ket{g}$ with no photonic component. Because of this, the memory generation must be heralded with a success probability in each shot being generally $p \sim 10^{-3}$. The undesired two-photon component scales as $p^{2}$ which can limit attempts to increase the rate of memory generation.

In the remainder of the paper, we will use a simplified notation where a bold-face letter, such as in, $\Ket{\bf{x}}$, represents a collective excitation in state $\Ket{x}$ that is in the form: 
\begin{equation}
\Ket{\bf{x}}=\frac{1}{\sqrt{N}}\sum_{j=1}^N e^{i\phi_j}\Ket{g,...x_j,...g} 
\label{eq:super2}
\end{equation}
Where `$x$' is any of the singly excited states (e.g. $ u$, $e$, $R$, etc.).

In order to ensure that the quantum memories do not de-phase during a storage time, it is desirable to transfer the single excitation from the Rydberg level, which has a relatively fast dephasing time, into a long-lived memory ground state $\Ket{\bf{u}}$, as in Fig.~\ref{fig:figone}. This is done by applying another two-photon $\pi$-pulse from $\Ket{\bf{R}}$ to $\Ket{\bf{u}}$ through $\Ket{\bf{e}}$, i.e. steps $\vec{\bf{ k_3}}$ and $\vec{\bf{k_4}}$ in Fig.~\ref{fig:figone} (a). The pulse sequence to generate a single memory state is shown in Fig.~\ref{fig:figone} (a). At this point, the atomic ensemble is in the memory state $\Ket{\bf{u}}$, (i.e. a collective excitation of the form in Eq.~(\ref{eq:super2})), with a phase ${\phi_j =(\vec{\bf{ k_1}}+\vec{\bf{k_2}}+\vec{\bf{k_3}}+\vec{\bf{k_4}})\cdot \vec{\bf{r_j}}}$. This process has recently been used to generate a Fock state of atoms by Ebert {\sl et al.}\cite{Ebert2014}.

To read the memory out photonically, we first apply a single-atom two-photon $\pi$-pulse from $\Ket{\bf{u}}$ to $\Ket{\bf{R}}$, steps $\vec{ \bf{k_5}}$ and $\vec{ \bf{ k_6}}$ in Fig.~\ref{fig:figone} (b). This is followed by applying strong blue light nearly resonant with the $\Ket{\bf{R}}$ to $\Ket{\bf{e}}$ transition, $\vec{\bf{k_7}}$. The state $\Ket{\bf{e}}$ quickly decays to the ground state. The amplitude of emission into a given spatial mode with associated wave vector $\vec{\bf{k_e}}$, is given by the condition: 
\begin{equation}
A\propto \frac{1}{N}\left|\sum_{j=1}^Ne^{-i({\vec{\bf{ k_{tot}}}-\vec {\bf{k_e}}})\cdot {\vec{ \bf{r_j}}}}\right|^2
\label{eq:collective}
\end{equation}
Where $\vec{\bf{k_{tot}}} = \sum_{j=1}^{7}(\vec{\bf{k_j}})$. In general, the spontaneous emission will be into $4\pi$, but emission into an arbitrary $\vec{\bf{k}}$ will result in the amplitude in Eq.~(\ref{eq:collective}) averaging to 1. In the particular case when the emitted photon ${\vec k_e}$ brings the excitation back to the reservoir state and ${\vec {\bf{k_e}}} = {\vec{ \bf{k_{tot}}}}$, the exponent in Eq.~(\ref{eq:collective}) is equal to zero, and the amplitude averages to $N$ \cite{Saffman2002}. This amounts to an enhancement of the spontaneous emission into a particular spatial mode, determined by the geometry of the pulses culminating in $\vec{\bf{k_{tot}}}$. Emission into the phase matched direction constructively interferes, while emission in an arbitrary direction destructively interferes \cite{Zeuthen2011}. In the presence of additional ground states, $\Ket{\bf{e}}$ decays preferentially to the reservoir state with an enhancement factor of $N$ \cite{Gorshkov2007}.

The process depicted in Fig.~\ref{fig:figone} of memory generation and photon retrieval can be viewed as an eight-wave mixing process analogous to viewing the DLCZ process as a coherent time-delayed four-wave mixing. This enhances the single-photon collection efficiency from the quantum memory.

Thus, the Rydberg blockade mechanism in an atomic ensemble can be used as a high-efficiency source of directionally emitted photons as was originally theoretically proposed by Lukin {\sl et al.}\cite{Lukin2001} and Saffman {\sl et al.}\cite{Saffman2002}. Collectively enhanced spontaneous emission of a single photon from a Rydberg media was demonstrated by Li {\sl et al.}\cite{Li2013}. 

Alternatively, one could read out the memory by exciting from $\ket{\bf{u}}$ directly to $\ket{\bf{e}}$ giving an effective six-wave mixing to extract a directional photon. However, in this case, the strong beam would be at nearly the same wavelength as the single-photon, whereas in the eight-wave mixing case, the strong de-excitation beam has a very large spectral difference from the single-photon (i.e. 480 nm vs. 780 nm in the case of $^{87}$Rb). Though filtering schemes have allowed for good readout for certain single-atom states \cite{Matsukevich2005}, our approach allows us the versatility to store multiple memories within one ensemble, hence necessitating more optical beams. The advantage for filtering the single-photon signal from the read pulse will likely outweigh the simplification of applying fewer beams, yet this remains a possibility to explore.

\section{\label{sec:level3.0}Experimental Realization}
\begin{figure}
\includegraphics[width=1.0\columnwidth]{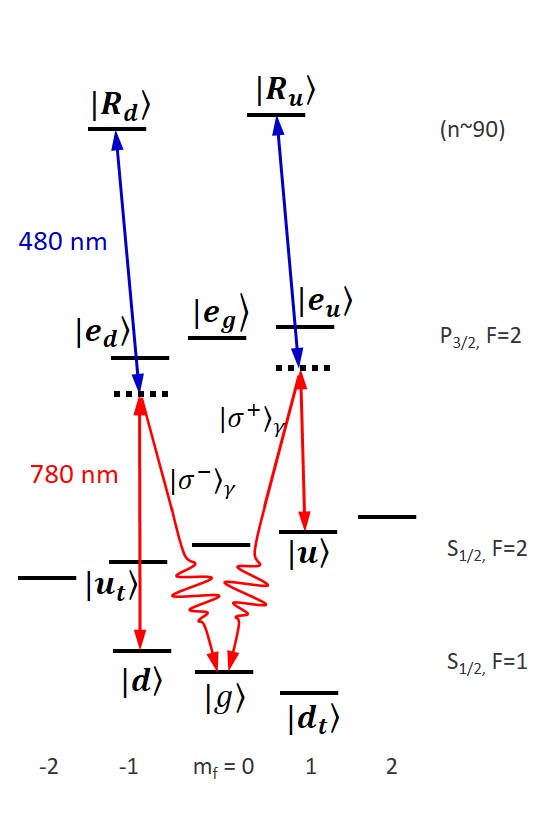}
\caption{\label{fig:figtwo}The energy level diagram of $^{87}$Rb used for the entanglement generation protocol where the bold indicates the light coupling to collectively excited states. The red light (780nm) couples the ground states ($\ket{\bf{u}}$ and $\ket{\bf{d}}$) to the intermediate excited states ($\ket{\bf{e_u}}$ and $\ket{\bf{e_d}}$). The blue light (480 nm) couples the intermediate excited state to two different Rydberg levels ($\ket{\bf{R_u}}$ and $\ket{\bf{R_d}}$). Light emitted when a state decays from the excited states $\ket{\bf{e_u}}$ or $\ket{\bf{e_d}}$ to $\ket{g}$ will have orthogonally polarized circular polarization, $\sigma^+$ or $\sigma^-$ respectively. The states with subscripts t are used later in the teleportation protocol}
\end{figure} 

We briefly discuss an experimental realization of these protocols using a laser cooled ensemble of $^{87}$Rb atoms. The relevant energy diagram of $^{87}$Rb is given Fig.~\ref{fig:figtwo}. The degeneracy of the hyperfine levels is lifted by applying a magnetic field of $\sim$ 0.5 mT which splits the ground states by $\sim3.5$ MHz.

We use a Rydberg level of n $\sim$90 which has been shown to have a lifetime of around $30$ $\mu s$ and $\sim$10 MHz Rydberg blockade shift at a distance of 10 $\mu$m with excitation laser linewidths on the order of a few kHz \cite{Johnson2008}. Hence we use 10 $\mu$m as $r_b$. Two-photon Rabi frequencies to n$\sim$90 Rydberg levels on the order of 1 MHz have been demonstrated with inferred fidelities of nearly $0.9$ \cite{Ebert2015}. The Rabi-frequency must be large enough to insure that operations can take place faster than the dephasing of the Rydberg levels. 

Multiple memories can be stored in the different Zeeman sublevels shown in Fig.~\ref{fig:figtwo}. For the entanglement generation protocol, we will use two ground-states, $\ket{\bf{u}}$ and $\ket{\bf{d}}$ for the quantum memory and a reservoir state $\ket{g}$. If beneficial for a particular protocol, the assignment of memory states and reservoir state could be changed. The states $\ket{\bf{u_t}}$ and $\ket{\bf{d_t}}$ will be used in Section~\ref{sec:level5}. The Zeeman-state coherence can be one limit to the memory lifetime. Maintaining control at the few percent level of a few mT magnetic field would still allow for $\sim 100$ $\mu s$ lifetime, which would be sufficient to link nodes at a distance of $\sim 20$ km. 

The memory states $\ket{\bf{u}}$ and $\ket{\bf{d}}$ are coupled to independently addressable Rydberg levels, $\ket{\bf{R_u}}$ and $\ket{\bf{R_d}}$, through different intermediate excited states $\ket{\bf{e_u}}$ and $\ket{\bf{e_d}}$. Photons emitted by a state decaying from $\ket{\bf{e_u}}$ or $\ket{\bf{e_d}}$ to $\ket{g}$ will have orthogonal circular polarizations, $\sigma^+$ or $\sigma^-$, respectively. These polarizations can be mapped onto the desired polarizations $\ket{V}$ or $\ket{H}$ with quarter wave-plates.

\begin{figure*}
\includegraphics[width=2.0\columnwidth]{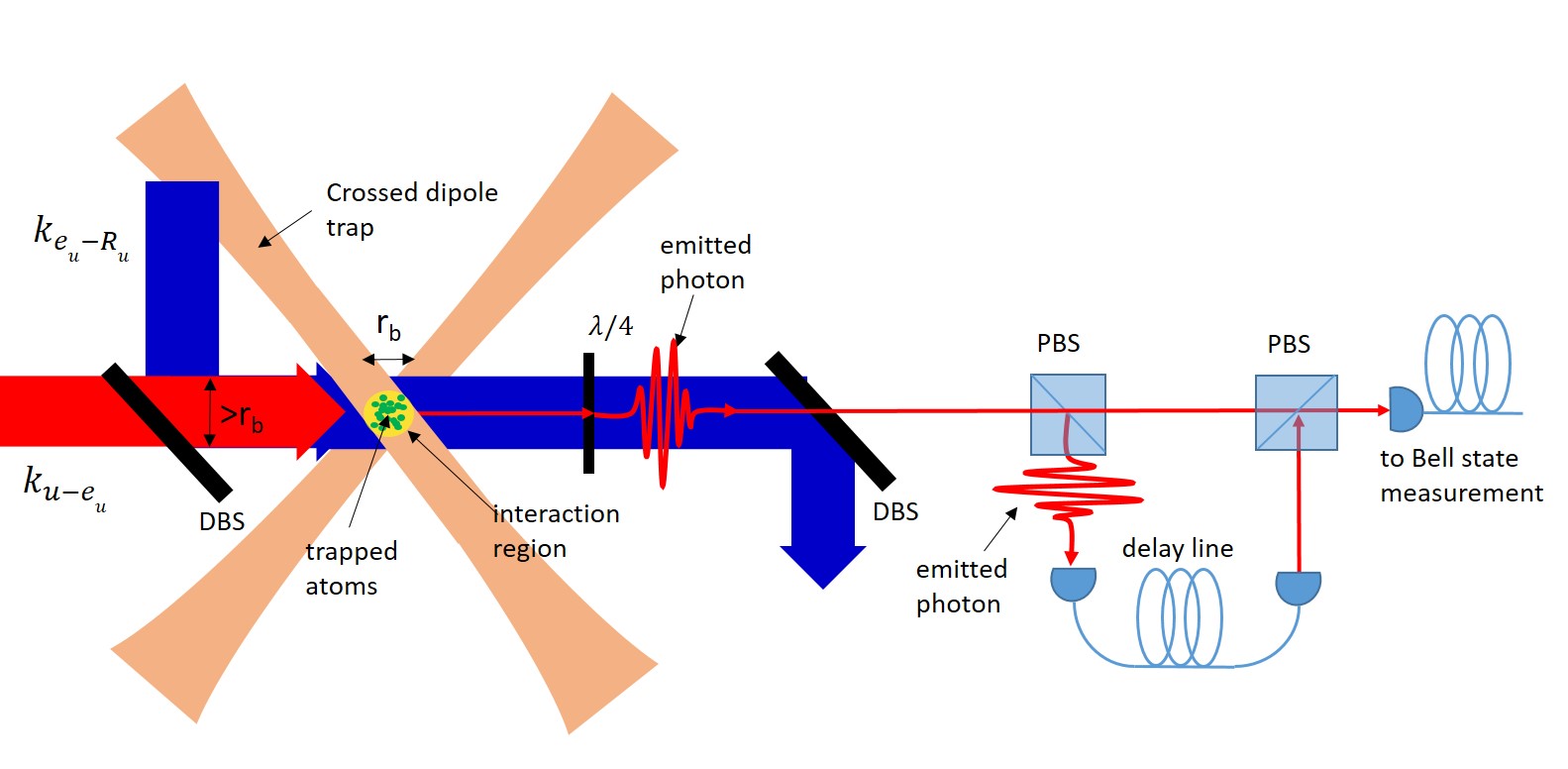}
\caption{\label{fig:excitation}The atoms are trapped in an optical dipole trap with a waist $< r_b$. The excitation beams ($k_{e_u-R_u}$ and $k_{u-e_u}$),  with a waist diameter $> r_b$, are overlapped on a dichroic beam splitter (DBS) and intersect the atoms perpendicularly to the trap. The excitation beams for the $\ket{d}$ states (not shown) can be overlapped with the excitation beams for the $\ket{\bf{u}}$ states using polarizing beam splitters. This produces a spherical interaction region (yellow) with a diameter $= r_b$ where a single excitation is produced. The single excitation is mapped onto a directionally emitted photon which can be converted into the desired polarization with a $\lambda/4$ wave-plate. A second DBS filters the blue excitation light out from the signal photon. The two outgoing photons can be overlapped in time with the use of a PBS and a delay line before being sent to a Bell state measurement}
\end{figure*}

The $^{87}$Rb atoms will be collected in a magneto-optical trap (MOT) and then loaded into a crossed optical dipole trap. If the optical trap has a diameter smaller than $r_b$, and the excitation beams have waists larger than $r_b$ we can ensure that the beams interact only with the atoms within one $r_b$, and that all atoms have the about same interaction strength with the excitation beams, see Fig.~\ref{fig:excitation}. We assume that the MOT temperatures is cold enough that it does not limit the memory lifetime. The memory lifetime can be increased by transferring the atoms to a far off-resonant optical lattice at the Rydberg magic wavelength \cite{Goldschmidt2015} or by using dynamic decoupling techniques \cite{Dudin2013}. Even without the use of a magic wavelength lattice, memory times of a few ms have been demonstrated in similar systems\cite{Ebert2015}. 

If quantum repeater nodes with Rydberg excitations are incorporated into a large scale fiber network, quantum frequency conversion would need to be used to overcome photon loss in the fiber. Frequency conversion of the rubidium signal in the near infrared (780 nm) to a telecom band ($\sim1324$ nm or $\sim1550$ nm) is promising as it can be done with a one-step conversion process. This could be implemented using the atoms as the non-linear device \cite{Radnaev2010}, or preferably, a non-linear crystal waveguide converter \cite{Albrecht2014}.

A full analysis of the potential errors and limiting factors is beyond the scope of this paper, but we list some of the more important ones here. Fluctuation of static electric and magnetic fields can interfere with the Rydberg or ground-state Zeeman levels. Atomic motion or collisions can lead to dephasing which can limit the single-photon collection efficiency and memory coherence time. Two-photon excitations or dark counts in single photon detectors can lead to erroneous heralding of entanglement generation. AC stark shifts from the trapping laser or excitation beams or off-resonant excitations to Rydberg levels can limit the state preparation fidelity.  In a dense atomic sample resonances with molecular Rydberg states can cause de-phasing\cite{Amthor2010}. In order to mitigate this the atoms can be loaded into a 3D optical lattice to control the inter-atomic spacing, and the principal quantum number and two-photon detuning can be adjusted to avoid the resonances. 

\section{\label{sec:level3}Generating a flying qubit entangled with memory}

\begin{figure}
\includegraphics[width=1\columnwidth]{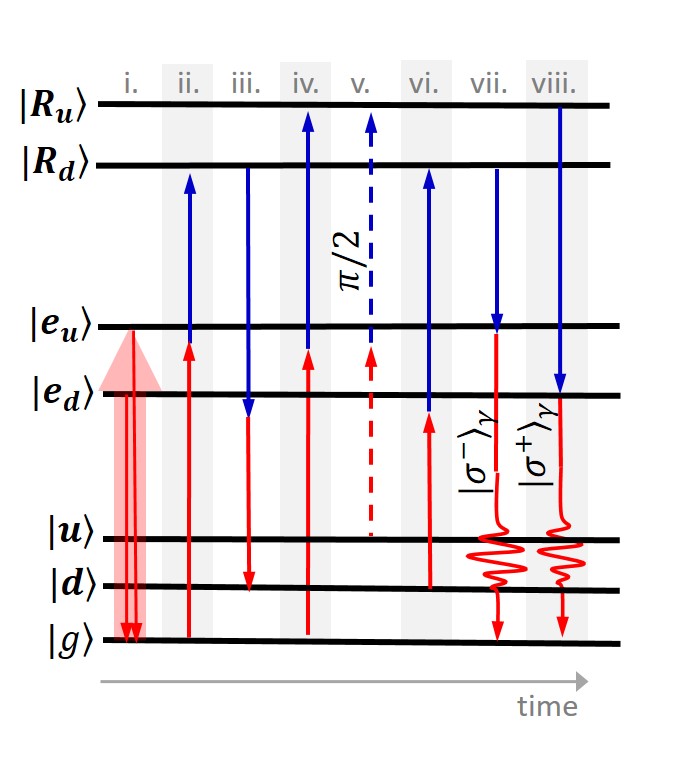}
\caption{\label{fig:entanglement}A graphical depiction of the steps in Table~\ref{tab:table1} using the state identified in Fig.~\ref{fig:figtwo}. Optical pumping in step i) is followed by a series of $\pi$-pulses (solid lines) and $\pi/2$-pulses (dashed lines) in order to prepare a memory qubit entangled with a flying photonic qubit. }
\end{figure}

Following the proposals of Zhao {\sl et al.}\cite{Zhao2010} and Han {\sl et al.} \cite{Han2010}, we describe how to prepare an entangled state suitable for quantum communication protocols. In short, we first produce two spin waves into different magnetic sublevels in the ensemble by applying the steps of Fig.~\ref{fig:figone} (a) twice, produce entanglement between them via a Rydberg blockade gate, and then read the components of that state into orthogonally polarized photons by applying the steps of Fig.~\ref{fig:figone} (b) twice. The state produced is comprised of two ground state memories entangled to photons of a flying polarization qubit. This structure, sometimes known as `dual-rail' \cite{Kok2007}, does not require interferometric, i.e. on the order of the wavelength, stability along the optical path \cite{Sangouard2011}. Rather it requires the path length to be stable to the level of the photon coherence length, which is much less stringent. The steps to produce this state are summarized in Table~\ref{tab:table1} and are detailed in the text, where all states are atomic states as labeled in Fig.~\ref{fig:figtwo} and the photonic states have as subscript $\gamma$. A graphical depiction is shown in Fig.~\ref{fig:entanglement} where the steps correspond to those in Table~\ref{tab:table1}. 

\begin{table}
\caption{\label{tab:table1}Entanglement preparation of flying qubit entangled with long-lived ground-state excitations, where $\pi_N$ identifies collectively enhanced rotations and $\pi$ identifies single-atom rotations.}
\begin{ruledtabular}
\begin{tabular}{l|l|l}
Step&Pulse&Result\\
\hline 
i.& Optically pump to ($\Ket{g}$) & $\Ket{g}$\\
ii. & $\pi_{N}$($\Ket{g}$ to $\Ket{\bf{R_{d}}}$) & $\Ket{\bf{R_{d}}}$\\ 
iii. & $\pi$($\Ket{\bf{R_{d}}}$ to $\Ket{\bf{d}}$) & $\Ket{\bf{d}}$\\
vi. & $\pi_{N}$($\Ket{g}$ to $\Ket{\bf{R_{u}}}$) & $\Ket{\bf{d}}\ket{\bf{ R_{u}}}$\\ 
v. & $\pi/2$($\Ket{\bf{R_{u}}}$ to $\Ket{\bf{u}}$) & $\Ket{\bf{d}}(\Ket{\bf{u}}+\Ket{\bf{R_{u}}})$\\
vi. & $\pi$($\Ket{\bf{d}}$ to $\Ket{\bf{R_{d}}}$)& $(\Ket{\bf{R_{d}}}\Ket{\bf{u}}+\Ket{\bf{d}}\Ket{\bf{R_{u}}})$\\
vii. & readout ($\Ket{\bf{R_{d}}}$ to $\Ket{\bf{e_{d}}}$) & $(\Ket{\sigma^-}_{\gamma}\ket{\bf{u}}+\Ket{\bf{d}}\ket{\bf{R_{u}}})$\\
viii. & readout ($\Ket{\bf{R_{u}}}$ to $\Ket{\bf{e_{u}}}$) & $(\Ket{\sigma^-}_{\gamma}\Ket{\bf{u}}+\Ket{\bf{d}}\ket{\sigma^+}_{\gamma})$\\
\end{tabular}
\end{ruledtabular}
\end{table}

The ensemble state is initialized in step i. of Table~\ref{tab:table1} by optically pumping all atoms to the reservoir state, $\ket{g}$. This can be done with $\pi$ polarized light if $\ket{g}$ is the $F=1, m_f =0$ state. The $N$ subscripts in Table~\ref{tab:table1} refer to transitions that have an enhanced effective Rabi frequency, all other pulses are for single atoms with single atom Rabi frequencies. All transitions between Rydberg levels and ground states are two-photon transitions. 

In step ii. of Table~\ref{tab:table1} we apply a two-photon $\pi_N$-pulse with an enhanced Rabi frequency from $\ket{g}$ to a high lying (n $\sim$ 90) Rydberg level, to create the state $\ket{\bf{R_d}}$. We use the intermediate excited state $F=2, m_f =-1$ of the $P_{3/2}$ D2 line, i.e. the state $\ket{\bf{e_{d}}}$. Next we shelve the Rydberg excitation in one of the ground states, $F=1, m_f =-1$, to produce the state $\ket{\bf{d}}$, by applying a $\pi$ pulse from $\ket{\bf{R_d}}$ through $\ket{\bf{e_{d}}}$, step iii. of Table~\ref{tab:table1}. Recall that here, and in the rest of the paper, all excited states are single excitation superpositions in the form of Eq.~(\ref{eq:super2}).

Next we excite a second Rydberg excitation,$\ket{\bf{R_u}}$, step iv of Table~\ref{tab:table1}, with a $\pi_{N}$ pulse from $\ket{g}$  which can be addressed independently from $\ket{\bf{R_d}}$ because of the frequency difference between $\ket{\bf{R_u}}$ and $\ket{\bf{R_d}}$. The state of the ensemble is now given by the product state of two ground state memories as shown at the end of step v. in Table~\ref{tab:table1}. The notation, $\ket{\bf{\bf{R_u}}}\ket{\bf{d}} \equiv \ket{\bf{\bf{R_u}d}}$  represents a double sum of product states analogous to Eq.~(\ref{eq:super2}) where each term in the sum is a product state of two different atoms in different states. In this way, multiple memories are stored within the same atomic ensemble \cite{Brion2007}. 

The Rydberg level $\ket{\bf{R_u}}$ is coupled to a second ground state memory $F=2, m_f =1$, $\ket{\bf{u}}$. The excitation to $\ket{\bf{u}}$ uses a different intermediate excited state, $F=2, m_f =1$ of the $P_{3/2}$ D2 line, $\ket{\bf{e_u}}$.

To produce an entangled state, in step v. of Table~\ref{tab:table1} we apply a $\pi/2$ pulse from $\ket{\bf{R_{u}}}$ to $\ket{\bf{u}}$ producing the superposition, see Fig.~\ref{fig:figtwo}:
\begin{equation}
1/\sqrt{2}\ket{\bf{d}}(\ket{\bf{u}}+\ket{\bf{R_{u}}})
\label{eq:psi1}
\end{equation}

This is followed by a $\pi$-pulse from $\ket{\bf{d}}$ to $\ket{\bf{R_{d}}}$, step vi. of Table~\ref{tab:table1}. Though $\ket{\bf{R_u}}$ and $\Ket{\bf{R_d}}$ are different, they still experience strong interactions and Rydberg blockade one another. In the $\ket{\bf{u}}\ket{\bf{d}}$ component $\ket{\bf{d}}$ is transferred to $\ket{\bf{R_d}}$, whereas in the $\ket{\bf{R_u}}\ket{\bf{d}}$ component, blockade between the Rydberg levels $\Ket{\bf{R_u}}$ and $\ket{\bf{R_d}}$ shifts the $\Ket{\bf{R_d}}$ state out of resonance, and the component is unchanged. The resulting state at the end of step vi. in Table~\ref{tab:table1} is given by:
\begin{equation}
\ket{\psi}=1/\sqrt{2}(\ket{\bf{u}}\ket{\bf{R_{d}}}+\ket{\bf{d}}\ket{\bf{R_{u}}})
\label{eq:psi}
\end{equation}
This is an entangled state between two Rydberg excitations and two ground state excitations in the same ensemble.

Next, a partial readout maps the components of the qubit into a photonic qubit. To map the excitation in $\ket{\bf{R_d}}$ to a photon, we apply a strong blue beam nearly resonant with the transition from $\ket{\bf{R_d}}$ to $\ket{\bf{e_d}}$, see step vii. of Table~\ref{tab:table1} and Fig.~\ref{fig:figtwo}. This intermediate excited state quickly decays to the reservoir state and preferentially scatters into the spatial mode set by the phase matching condition for the eight-wave mixing process as discussed previously in relation to Eq.~(\ref{eq:collective}). This is followed by mapping the $\ket{\bf{R_u}}$ state onto a photon with a beam nearly resonant with the $\ket{\bf{R_u}}$ to $\ket{\bf{e_u}}$ transition, step viii. of Table~\ref{tab:table1}. The two memories could potentially be read simultaneously, which would reduce the time it takes to perform the atomic protocol, though not the probability of successfully reading out the state, as discussed in Section ~\ref{sec:level5.5}.

Alternatively the qubits could be read sequentially, in which case the time-bin qubit can be mapped onto a polarization qubit by delaying the first with a delay line, such as a long optical fiber and then overlapping it with the second photon, as shown in Fig.~\ref{fig:excitation}. A fiber delay of a few hundred meters will be sufficient to overlap the two photons but will not significantly contribute to the distance the photon must travel in a fiber to entangle two nodes located several $\sim$ km apart. 

The photons read from $\ket{\bf{u}}$ and $\ket{\bf{d}}$ have orthogonal circular polarization and we have arrived at the end of step ix. of Table~\ref{tab:table1}. As shown in Fig.~\ref{fig:entanglement}, using a $\lambda/4$ wave-plate, we rotate the $\ket{\sigma^-}_{\gamma}$ and $\ket{\sigma^+}_{\gamma}$ photons into horizontal and vertical polarization respectively, which we label $\ket{H}_{\gamma}$ and $\ket{V}_{\gamma}$ to obtain the state:
\begin{equation}
\ket{\psi}=1/\sqrt{2}(\ket{\bf{u}}\ket{H}_{\gamma}+\ket{\bf{d}}\ket{V}_{\gamma})
\label{eq:mem}
\end{equation}

Here we have a maximally entangled state between a flying photonic polarization qubit and long-lived ground state memories. Importantly, because the two photons are emitted with orthogonal polarizations and the single photon is emitted hundreds of nm detuned from the de-excitation beam it should be possible to implement this protocol with all beams on a single axis combined with dichroic beam splitters and polarizing beam splitters as shown in Fig.~\ref{fig:excitation}. The dichroic beam splitter downstream from the ensemble filters the de-excitation beam from the signal photon. Additional filters will likely be needed to fully attenuate the de-excitation beam from the single photon signal.

This is in contrast to the DLCZ schemes where off-axis collection of the single photon is extremely useful to aid in filtering out the de-excitation beam and the close spectral proximity of the signal to the pump can require spectral filtering \cite{Stack2015}. Further, off-axis geometry can limit the memory lifetime \cite{Zhao2009}. The switch to an on-axis geometry greatly reduces the experimental alignment complexity and should aid in achieving a high memory read-out efficiency. 

The state in Eq.~\ref{eq:mem} is suitable for many quantum communication protocols including repeaters, as described in Section~\ref{sec:level4}, or teleportation as described in Section~\ref{sec:level5}.


Our protocol differs from the Zhao {\sl et al}.\cite{Zhao2010} proposal by skipping several steps. We read the Rydberg states out directly, whereas they shelve the Rydberg excitations to two additional ground state levels for long-term memory and deterministic on-site entanglement swapping. However, in the case of two-node communication this is unnecessary and we can simply read the state out directly. This enables us to produce entangled flying qubits with fewer steps and use three ground-states instead of five. This reduces experimental complexity for two-node protocols at the expense of the many node scaling of the full protocol, as will be described in the following section. 

\section{\label{sec:level4}Generating remote entanglement}

To generate entanglement between two different remote ensembles consider two systems, A and B each prepared in a state in the form of Eq.~(\ref{eq:mem}). The resulting combined state is:
\begin{equation}
1/2(\ket{ \bf{u_{A}}}\Ket{H_{A}}_{\gamma} + \Ket{\bf{d_{A}}}\ket{ V_{A}}_{\gamma})\otimes (\ket{\bf{ u_{B}}}\Ket{H_{B}}_{\gamma} + \Ket{\bf{d_{B}}}\ket{ V_{B}}_{\gamma})
\label{eq:generation}
\end{equation}

\begin{figure}
\includegraphics[width=1.0\columnwidth]{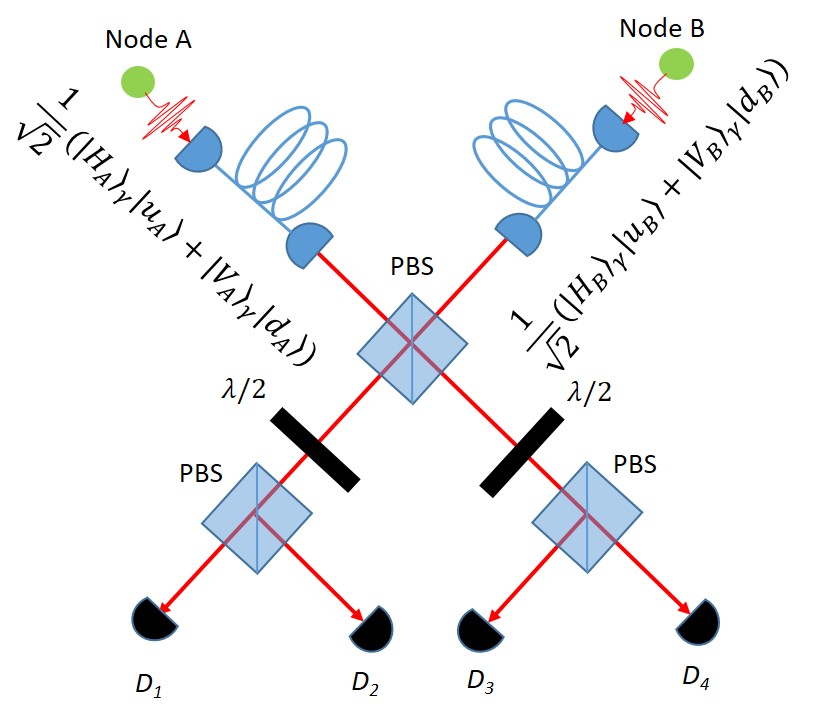}
\caption{\label{fig:bell} Photons from the nodes A and B are input to the Bell state analyzer and are interfered on a PBS with axes orthogonal to the polarization of light. Then each arm is then sent through a $\lambda/2$ wave-plate which rotates the polarization by 45 degrees. The photons in each arm are then sent through another PBS. All four output ports are measured with single photon detectors, $D_1$-$D_4$.}
\end{figure} 

The flying qubits from A and B are overlapped on a polarizing beam splitter (PBS) and then subject to a Bell state analyzer\cite{Pan1998} such as shown in Fig.~\ref{fig:bell}. The axes of the PBS are oriented in the H and V axis of the light. The two outputs of the PBS are sent through $\lambda/2$ wave-plates which rotate the polarization by 45 degrees and are then incident on a second PBS. All four output ports are measured with single photon detectors ($D_1 - D_4$). Particular pairs of two photon coincidence measurements will project the state onto a Bell state. In this setup, if coincidences counts between $D_1$ and $D_4$ or $D_2$ and $D_3$ are measured, we project the remaining wave-function onto the state $1/\sqrt{2}( \ket{\bf{u_{A}}}\ket{\bf{u_{B}}} + \ket{\bf{d_{A}}}\ket{\bf{d_{B}}} )$. However, if on the other hand we measure coincidence counts between $D_1$ and $D_3$ or $D_2$ and $D_4$, then we produce the state $1/\sqrt{2}( \ket{\bf{u_{A}}}\ket{\bf{u_{B}}} - \ket{\bf{d_{A}}}\ket{\bf{d_{B}}} )$. In the remaining half of the terms in the expansion of Eq.~(\ref{eq:generation}), two photons are sent to one detector, and since we cannot discriminate photon number these counts are lost. Thus if coincidence counts between $D_1$ and $D_4$ or $D_2$ and $D_3$ are measured we produce the desired state. If coincidence counts between $D_1$ and $D_3$ or $D_2$ and $D_4$ are measured, then we perform a local unitary operation to transform the state into the desired state:
\begin{equation}
\ket{\psi}=1/\sqrt{2}( \ket{\bf{u_{A}}}\ket{\bf{u_{B}}} + \ket{\bf{d_{A}}}\ket{\bf{d_{B}}} )
\label{eq:bell}
\end{equation}
This state represents a maximally entangled state between two remote long-lived ground state memories and can be used as the base entanglement resource for further protocols.

\begin{figure}
\includegraphics[width=1.0\columnwidth]{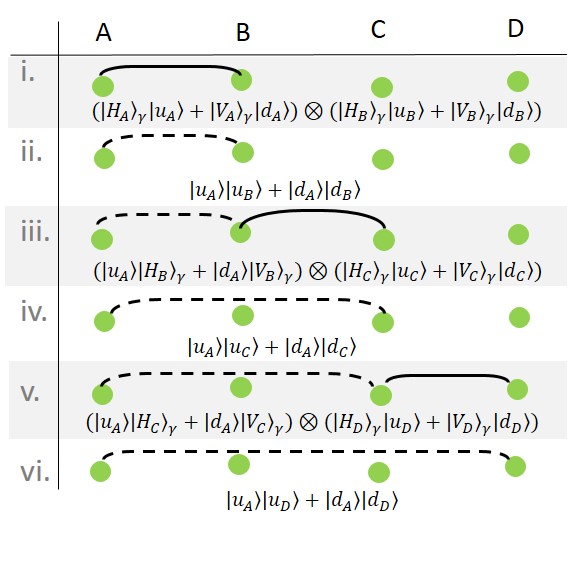}
\caption{\label{fig:distribution} Graphical depiction of entanglement distribution. The nodes are labeled A-D. A solid line represents photons being passed between two nodes. A dashed line represents entanglement established between two nodes. i) The ensembles A and B are prepared in the state described by Eq.~\ref{eq:bell}. ii) A and B are entangled via a Bell state measurement producing the state in Eq.~\ref{eq:mem}. iii) The qubits in B and C are mapped onto photons. iv) A Bell state measurement between the photons from B and C extends the entanglement from A to C. v) The qubits in C and D are mapped onto photons. vi) A Bell state measurement between the photons from C and D extends the entanglement from A to D.}
\end{figure}

To use this protocol in a repeater to distribute entanglement we take Eq.~\ref{eq:bell} as the starting point and iterate the procedure in Table~\ref{tab:table1}. This simple model captures several of the features of many-node networks. Fig.~\ref{fig:distribution} shows the sequence of entanglement distribution. First we prepare A and B in the state described by Eq.~\ref{eq:bell}, as seen in step i. of Fig.~\ref{fig:distribution}. Next, system A and B are entangled via a Bell state measurement as described above and depicted in step ii. of Fig.~\ref{fig:distribution}. Then, we simultaneously map the qubit in B onto a photonic qubit and prepare ensemble C in the state $1/\sqrt{2}(\ket{ \bf{u_{C}}} \Ket{H_{C}}_{\gamma}+ \Ket{\bf{d_{C}}}\ket{ V_{C}}_{\gamma})$. This results in the composite state shown in Eq.~\ref{eq:bell2}.

\begin{equation}
1/2( \ket{\bf{u_{A}}}\ket{H_{B}}_{\gamma} + \ket{\bf{d_{A}}}\ket{V_{B}}_{\gamma} ) \otimes(\ket{ \bf{u_{C}}}\Ket{H_{C}}_{\gamma} + \Ket{\bf{d_{C}}}\ket{ V_{C}}_{\gamma})
\label{eq:bell2}
\end{equation}

Next, the flying qubits from system B and C are entangled via a Bell state measurement, step iii. in Fig.~\ref{fig:distribution}. The entanglement between A and B is then transferred to entanglement between A and C, and the result is the long-lived entangled state given in step iv and Eq.~\ref{eq:bell3}:
\begin{equation}
1/\sqrt{2}( \ket{\bf{u_{A}}}\ket{\bf{u_{C}}} + \ket{\bf{d_{A}}}\ket{\bf{d_{C}}} )
\label{eq:bell3}
\end{equation}
This process can be repeated to continue spreading the entanglement to further nodes as long as the memories in A or any node to which A is entangled do not decohere. This could require the use of high-speed optical switches to ensure that the emitted photons are directed towards the appropriate links. 

\begin{center}
\begin{table*}\caption{\label{tab:table2}Teleportation of a collective excitation from one ensemble to a remote ensemble}
\begin{ruledtabular}
\begin{tabular}{l|l|c}
&Pulse&Result\\
\hline
i. & initial & $\alpha(\Ket{\bf{uuu}} + \Ket{\bf{udd}}) + \beta (\ket{\bf{duu}} + \Ket{\bf{ddd}})$ \\ 
ii. & $\pi$($\Ket{\bf{u_{t}}}$ to $\Ket{\bf{R_{d}}}$) & $\alpha (\Ket{\bf{Ruu}} + \Ket{\bf{Rdd}}) + \beta (\Ket{\bf{duu}} + \Ket{\bf{ddd}})$ \\ 
iii. & $\pi/2$($\Ket{\bf{u_{A}}}$ to $\Ket{\bf{d_{A}}}$) & $\alpha/\sqrt{2}(\Ket{\bf{R}(\bf{u}+\bf{d})\bf{u}} + \Ket{\bf{R}(\bf{u}-\bf{d})\bf{d}}) + \beta/\sqrt{2}(\Ket{\bf{d}(\bf{u}+\bf{d})\bf{u}} + \Ket{\bf{d}(\bf{u}-\bf{d})\bf{d}})$ \\
iv. & $2\pi$($\Ket{\bf{u_{A}}}$ to $\Ket{\bf{R_{u}}}$) & $\alpha/\sqrt{2} (\Ket{\bf{R}(\bf{u}+\bf{d})\bf{u}} + \Ket{\bf{R}(\bf{u}-\bf{d})\bf{d}}) - \beta/\sqrt{2}(\Ket{\bf{d}(\bf{u}-\bf{d})\bf{u}} + \Ket{\bf{d}(\bf{u}+\bf{d})\bf{d}})$ \\ 
v. & $\pi/2$($\Ket{\bf{u_{A}}}$ to $\Ket{\bf{d_{A}}}$) & $\alpha(\Ket{\bf{Ruu}} + \Ket{\bf{Rdd}}) - \beta(\Ket{\bf{ddu}} + \Ket{\bf{dud}})$ \\ 
vi. & $\pi$($\Ket{\bf{R_{d}}}$ to $\Ket{\bf{u_{t}}}$) & $\alpha(\Ket{\bf{uuu}} + \Ket{\bf{udd}}) + \beta(\Ket{\bf{ddu}} + \Ket{\bf{dud}})$ \\ 
\end{tabular}
\end{ruledtabular}
\end{table*}
\end{center}

We note that we could use this type of protocol to perform the nested entanglement generation and entanglement swapping architectures that are characteristic of repeater protocols\cite{Briegel1998}. For example, to entangle nodes A and D we could first produce the state:

\begin{equation}
1/2( \ket{\bf{u_{A}}}\ket{\bf{u_{B}}} + \ket{\bf{d_{A}}}\ket{\bf{d_{B}}})\otimes( \ket{\bf{u_{C}}}\ket{\bf{u_{D}}} + \ket{\bf{d_{C}}}\ket{\bf{d_{D}}})
\label{eq:bell4}
\end{equation}

Where we could wait until the entanglement between the pairs AB and CD are both successful. We could swap entanglement by reading out the photons in B and C and entangling them on a Bell state analyzer which would produce the state: 
\begin{equation}
1/\sqrt{2}( \ket{\bf{u_{A}}}\ket{\bf{u_{D}}} + \ket{\bf{d_{A}}}\ket{\bf{d_{D}}})
\label{eq:bell5}
\end{equation}

This approach relies on photon detection for the entanglement swapping and does not take advantage of the deterministic entanglement swapping proposed by the earlier protocols. Deterministic entanglement swapping relies on the use of additional ensembles at each node such as the approach in Zhao {\sl et al.} \cite{Zhao2010} which would require additional atom traps or makes more use of the multi-plexed quantum memory storage in the multiple ground states of an ensemble such as in Han {\sl et al.} \cite{Han2010} which would require more addressing laser beams. However, since we are more concerned with comparing the performance of entanglement generation and teleportation for two-nodes, we take advantage of the reduction in experimental complexity and analyze the performance of the protocol using one ensemble and a minimum number of ground state levels. The experimental implementation we have described could be adapted to include the deterministic entanglement swapping as described by earlier proposals.

The errors that could arise in this protocol are primarily given by the two-photon errors and the detector dark counts. The two-photon error arises when two Rydberg excitations are produced in one ensemble at the same time. This is dependent on the Rydberg blockade detuning, laser linewidth, and off-resonant coupling strength. If two photons are produced, the additional photon could trigger a detector which would mistakenly herald the creation of an entangled state, and would thus produce an error. Similarly, a dark count would mistakenly herald the creation of a memory when none had actually been created. More detailed analysis of errors in these types of systems are given by Zhao {\sl et al.} \cite{Zhao2010} and Han {\sl et al.} \cite{Han2010}.

\section{\label{sec:level5}Teleportation between remote ensembles}

We wish to teleport an unknown quantum state from node A to node B. To do this we need two additional ground states which will store the state that will be teleported. We must also perform qubit rotations which could be done with a Raman transition between Zeeman sublevels\cite{Saffman2005}. We choose the $F=2, m_f=-1$ and $F=1, m_f=1$ ground states as the qubit encoding the state to be teleported, see Fig.~\ref{fig:figtwo}. These states will be used to produce the states $\ket{\bf{u_t}}$ and $\ket{\bf{d_t}}$. This `target' qubit pair is only used at node A, not at node B.

The $\ket{\bf{u_t}}$ and $\ket{\bf{d_t}}$ states share the intermediate excited states and Rydberg levels associated with the $\ket{\bf{d}}$ and $\ket{\bf{u}}$ states respectively. Making this choice allows us to access all the necessary states while minimizing the number of Rydberg excitations, reducing experimental complexity. Because two-qubit gate operations only occur between the original pairs or the target pairs, re-using the Rydberg states for the target pairs will not compromise the protocol even though the initial qubit at node A and the target qubit are physically located in the same ensemble of atoms. 

The two remote ensembles are initially prepared in the state in Eq.~\ref{eq:bell}. We establish entanglement by using the steps described in Section ~\ref{sec:level3} and Section ~\ref{sec:level4}. This generates the entanglement resource shared between two remote ensembles. Next, we produce the target state by exciting a spin wave in ensemble A to the state $\ket{\bf{u_t}}$ analogously to how we produced the state $\ket{\bf{u}}$. This results in the state $1/\sqrt{2}\ket{\bf{u_t}}(\Ket{\bf{u_{A}u_{B}}} + \Ket{\bf{d_{A} d_{B}}})$. This is followed by an off-resonant two-photon Raman transition from $\ket{\bf{u_t}}$ to $\ket{\bf{d_t}}$ through an intermediate excited state $F=2, m_{F}=0$, $\ket{\bf{e_{g}}}$. The Raman pulse can be chosen to give any arbitrary rotation, resulting in the state $\alpha \ket{\bf{u_t}} + \beta \ket{\bf{d_t}}$. The energy levels used in the teleportation protocol are shown in Fig.~\ref{fig:figtwo}, but the beams are not.

After generating entanglement between the two systems and producing the target state, the wave-function of the system is described by:
\begin{equation}
\ket{\psi}=1/\sqrt{2}(\alpha \Ket{\bf{u_t}} + \beta \Ket{\bf{d_t}})\otimes (\Ket{\bf{u_{A}u_{B}}} + \Ket{\bf{d_{A} d_{B}}})
\label{eq:six}
\end{equation}
The steps to perform teleportation between nodes A and B are summarized in Table~\ref{tab:table2}. The notation is simplified by identifying the first, second, and third elements in a ket as belonging to the `target' qubit, the initial A qubit, and the B qubit respectively.

After the initial state, we transfer the components of the target state that are in $\ket{\bf{u_t}}$ to the Rydberg state $\ket{\bf{R_d}}$ with a two-photon $\pi$-pulse, step ii. in Table~\ref{tab:table2}. This effectively blocks any transitions in the $\alpha$ component of the wave-function until the end of the protocol. Subsequent operations only affect components in the $\beta$ component of the wave-function. 

Next, in step iii of Table~\ref{tab:table2}, we perform a $\pi/2$ pulse between $\ket{\bf{u_A}}$ and $\ket{\bf{d_A}}$ with an off-resonant two-photon Raman transition. This transfers $\ket{\bf{u_A}}$ to $1/\sqrt{2}(\ket{\bf{u_A}} + \ket{\bf{d_A}})$ and $\ket{\bf{d_A}}$ to $1/\sqrt{2}(\ket{\bf{u_A}} - \ket{\bf{d_A}})$.

In step iv. of Table~\ref{tab:table2} we apply a $2\pi$ pulse from $\ket{\bf{u_A}}$ to $\ket{\bf{R_u}}$. Again, since the $\alpha$ component of the state already contains a Rydberg excitation, any state transfer on these states is prohibited by Rydberg blockade. In the $\beta$ component, the state $\ket{\bf{u_A}}$ receives a $\pi$ phase shift, i.e. it acquires a negative sign.

Next, in step v. in Table~\ref{tab:table2}, we perform another $\pi/2$ rotation between $\ket{\bf{u_A}}$ and $\ket{\bf{d_A}}$. This results in the A qubit on the $\beta$ component swapped with respect to the initial state. Finally, in step vi. of Table~\ref{tab:table2}, the state $\ket{\bf{R_d}}$ is rotated back to $\ket{\bf{u_t}}$, which acquires a $\pi$ phase shift because it has accumulated a $2\pi$ rotation. The result is the final state in Table~\ref{tab:table2} which has an overall $\pi$ phase shift removed. Table~\ref{tab:table2} essentially performs a CNOT gate with the `target' qubit as the control and the A qubit as the target \cite{Isenhower2010, Maller2015}

At the end of Table~\ref{tab:table2}, we now rotate the `target' qubit using a $\pi/2$ Raman pulse. If we apply this to the result of Table~\ref{tab:table2} and rearrange the terms, the total state of the system is now given by:

\begin{eqnarray}
\ket{\psi} = & 1/2\ket{\bf{uu}}(\alpha \ket{\bf{u}}+ \beta \ket{\bf{d}}) \nonumber \\ 
&+1/2\ket{\bf{du}}(\alpha \ket{\bf{u}}- \beta \ket{\bf{d}}) \nonumber \\ 
&+1/2\ket{\bf{ud}}(\alpha \ket{\bf{d}}+\beta \ket{\bf{u}}) \nonumber \\ 
&+1/2\ket{\bf{dd}}(\alpha \ket{\bf{d}}-\beta \ket{\bf{u}})
\label{eq:measure}
\end{eqnarray}
From this state it is clear that if the two qubits at A are measured, this projects the state of the B qubit onto one of the terms in Eq.~(\ref{eq:measure}). Once the classical result of the measurement is sent from A to B, then B can perform the appropriate rotation on the state of the qubit in B to produce the initially desired target state $\alpha \ket{\bf{u}}+ \beta \ket{\bf{d}}$.

\section{\label{sec:level5.5}Two node rate analysis}

To estimate the average time required for any protocol to be successful, we compute the sum of the time required for each step of the protocol, $t_i$, divided by the product of probabilities of success for each step, $p_i$:

\begin{equation}
T=\frac {\sum t_{i}}{\prod p_{i}}
\label{eq:seven}
\end{equation}

A quantum network with separated nodes has a round-trip time of light between the two nodes ($2d/c$), where $d$ is the optical distance between nodes and $c$ is the speed of light. One factor of $d/c$ accounts for the time of transmission of the flying qubit from one node to the second, and the other $d/c$ accounts for the time it takes the result of the measurement to be transmitted back to the first node as classical information. As the distance between nodes increases the total time of the protocol can become dominated by the round trip time of the light. For example, for a $\sim$10 km node separation, the time of flight, $d/c$, is 50 $\mu$s in an optical fiber. The round trip travel time of light sets an absolute maximum entanglement generation rate assuming there are no losses and that the atomic protocol is significantly faster than the light travel time. For 10 km the speed of light sets an absolute maximum rate of entanglement generation 10 kHz in an optical fiber. 

However, for estimating protocol times for short distance light propagation, we must look in more detail at the time it takes to perform the atomic protocol. The transition from a ground state to a Rydberg state is repeated many times for the protocol. For this estimate we ignore factors of two for $\pi/2$ pulses or $\sqrt{N}$ for the collective enhancement and assume that all of these are given by the rate of a two-photon $\pi$-pulse determined by the two-photon Rabi frequency, $\Omega_{R}$. Since a relatively small proportion of the pulses are $\pi/2$ pulses or have an Rabi frequency enhanced by $\sqrt{N}$, this will not significantly change the result. In similar experimental arrangements to the proposed one, i.e. $n=90$ Rydberg levels excited in $^{87}$Rb, two-photon Rabi frequencies as high as 750 kHz were achieved\cite{Li2013, Ebert2014}. In order to make calculations simple, we estimate the time of a single operation to be $t_o =1 \mu$s. A summary of the definitions of the parameters and subscripts used in Sections ~\ref{sec:level5.5} and ~\ref{sec:level6} is given in Table ~\ref{tab:table3}.

Interestingly, a potential source of additional time to perform the atomic protocol is the time it takes the photon to exit the cloud, which depends on the group velocity and ensemble size. The group velocity of light in the very similar experimental arrangement, i.e. an electromagnetically-induced-transparency beam configuration using Rydberg excitations in rubidium, has been shown to be in the range of 10 to 30 m/s \cite{He2014}, which, though very slow for light, will leave the 10 $\mu$m blockade radius in 0.1 to 0.3 $\mu$s. For this estimate, we take advantage of the comparable time scales and assume the readout time is close enough to the operation time, $t_o =1 \mu$s. With the simplifications above, we estimate the total time of the protocol to generate entanglement between two nodes with the total number of operations required, $n_{G}$, and the distance between nodes:
\begin{equation}
t_G=n_G t_o + 2d/c
\label{eq:rate1}
\end{equation}

The total number of transitions to and from the Rydberg level and photon readout for the entanglement generation is $n_G=7$, as read off Table~\ref{tab:table1}. Since the two ensembles can be prepared simultaneously, we do not need to include a factor of two in the atomic protocol time. If we assume the nodes are separated by a minimal distance, the time of the atomic protocol $n_G t_o$, sets the absolute maximum repetition rate of the experiment to be 140 kHz. 

More critical for the average time estimate of Eq.~\ref{eq:seven} are the success probabilities. The four-photon transitions through a Rydberg level and down to atomic ground states have been performed with a probability of $0.62$ \cite{Ebert2014}. Thus, we estimate the probability of success for a single transition to the Rydberg level to be $P_R=\sqrt{0.62}=0.79$. Given an atomic density, the photon collection efficiency is estimated by\cite{Gorshkov2007}:
\begin{equation}
p_{\gamma}=1-\frac{\sqrt{4/\pi}}{\sqrt{OD}}
\label{eq:rate}
\end{equation}

An atomic density of $n=5 \times 10^{11} cm^{-3}$ and $r_b$ of $10 \mu$m leads to $\sim 2000$ atoms within the blockade radius and an optical density (OD) per blockade radius of 3 which predicts a photon collection efficiency of $P_{\gamma} = 0.3$. This photon collection efficiency is typical in neutral atom ensemble memories without Rydberg excitations\cite{Stack2015}. A photon collection efficiency 11 \% has already been achieved for the first attempt of collecting photons from a collective Rydberg excitation in an ensemble\cite{Li2013} which had a comparable atom number within the blockade radius. Shelving a single Rydberg excitation into a long-lived ground state with the higher atom numbers needed for high photon collection efficiency has not yet been experimentally realized. In order to achieve reasonable fidelities, 3D optical lattices can be employed and different principal quantum numbers and detunings can be used to avoid de-phasing resonances. In addition, there is no fundamental reason that the collection efficiency from a memory produced via Rydberg excitations would be smaller than  from other neutral atom memories. Finally, we have a probability of obtaining a useful Bell state from the Bell state analyzer of $P_{B}=1/2$. 


\begin{table}
\caption{\label{tab:table3}Summary of parameters and subscripts used in Sections ~\ref{sec:level5.5} and ~\ref{sec:level6} }
\begin{ruledtabular}
\begin{tabular}{l|l}
Parameter&Description\\
\hline 
$n$ & Number of nodes\\
$n_G, n_T, n_S$ & Number of steps in a protocol\\
$t_o, t_G, t_T$ & Time for an single protocol\\ 
$P_\gamma, P_R,P_B,P_G,P_T$ & Probability of success\\
$T_G,T_T$ & Average time for a single protocol\\
$T_G[n],T_T[n],T_{T'}[n]$ & Average time for an n step protocol\\
\hline
\hline
Subscript & Description \\
\hline
$G$ & Entanglement generation \\
$S$ & Extending entanglement\\
$T$ & Teleportation protocol\\
$T'$ & Alternative protocol (see Fig.~\ref{fig:tele})\\
$o$ & Single operation \\
$\gamma$ & Photon collection \\ 
$R$ & Rydberg excitation efficiency\\
$B$ & Bell measurement\\
\end{tabular}
\end{ruledtabular}
\end{table}


Thus, we estimate the average time to successfully generate remote entanglement between two nodes as: 
\begin{equation}
T_G=\frac{(n_G-1) t_o + 2d/c}{(P_{R}^{n_{G}}P_{\gamma}^{2})^{2}P_{B}}=\frac{t_G}{P_G}
\label{eq:rate2}
\end{equation}

The factor $(P_{R}^{n_{G}}P_{\gamma}^{2})$ in the denominator is squared because of the fact that both ensembles must produce a flying qubit entangled with the memory. There is one Bell-state measurement. The factor of $(n_G-1)$ appears in the numerator because we have assumed simultaneous readout, whereas the denominator contains $n_G$ because simultaneous readout doesn't change the probability of successfully performing the operations. We label the denominator, $P_G$, as the probability of successfully generating entanglement and the numerator, $t_G$ as the total time for the atomic protocol and light travel time.

We use values that have been observed in experiments as identified above, $P_R =0.79$, $P_\gamma=0.3$, $t_o=1$ $\mu$s, and $P_B$. We also set $d=0$, which allows us to compare with other teleportation protocols over small distances. We calculate the average rate of entanglement generation using Eq.~\ref{eq:rate2} to be $1/T_G=25 $Hz. This is the same order of magnitude as the highest rates between matter qubits currently reported in ion entanglement at $\sim 5$ Hz\cite{Hucul2015}. 

Similarly, we can estimate the total time it takes to successfully perform teleportation by using Eq.~\ref{eq:seven}. We use entanglement generation as the first step in the teleportation protocol. Since the entanglement generation is heralded, if we use the average time to generate entanglement $T_G$ given in Eq.~\ref{eq:rate2}, we do not need to include the probability $P_G$ in the estimate for average teleportation time, as it is included in the estimate of $T_G$.

Thus, the estimate for the average time to successfully teleport a quantum state between two remote nodes is: 
\begin{equation}
T_T=\frac{T_G+n_{T}t_o+ 2d/c}{P_{R}^{n_{T}}}=\frac{T_G+t_T}{P_T}
\label{eq:rate3}
\end{equation}

Where $n_T$ is the total number of operations used for the teleportation protocol, i.e. the steps in Table~\ref{tab:table2}, as well as a pulse to project the final state onto the desired state given the result of the Bell-state measurement. For simplicity in this estimate we take $n_T \simeq n_G=7$. The time of the teleportation protocol is defined as $t_T=n_{T}t_o+ 2d/c$. The final state measurement can in principle be done with near unit efficiency with field selective ionization \cite{Rosenfeld2009} and is not considered for this estimate. The total probability for the steps used in just the teleportation protocol, $P_T$ thus does not include any photon collection, which again, was included in the estimate of $T_G$. 
Using the same values as above for the efficiency parameters, we estimate the rate of successful teleportation events predicted by Eq.~\ref{eq:rate3} to be $\sim$ 5 Hz. A $\sim$Hz rate is on the same order as the rate of teleportation achieved with a single-atom coupled to a high-finesse optical cavity \cite{Nolleke2013} whereas typical realizations of teleportation between matter qubits have had rates on the order of one every few minutes\cite{Olmschenk2009, Bao2012, Pfaff2014}. 

To improve the rate of entanglement generation, the transition efficiency to the collective states and the photon collection efficiency need to be improved. To see the effect of higher efficiencies on the protocol rates, we estimate the efficiencies that might be achieved with improved technology. One potential for improving the Rydberg transition probability is to use techniques such as the pulse shaping developed by Beterov {\sl et al.}\cite{Beterov2013}. In addition, the atoms could instead be loaded into a three-dimensional optical lattice in order to fix the separation between pairs of atoms and eliminate their motion. For this estimate, we use an improved Rydberg transition efficiency of $P_R = 0.9$ which is consistent with the theoretical prediction in Ref. \cite{Ebert2014} for higher atom numbers. To dramatically improve the photon collection efficiency, the ensemble can be coupled to a high finesse optical cavity. Photon collection efficiency as high as 0.84 has been achieved in non-Rydberg ensemble based systems \cite{Simon2007}. For this estimate we use a photon collection efficiency of $P_{\gamma}=0.80$. If these efficiencies can be achieved, these improved parameters would predict average rates of success for entanglement generation and teleportation given by Eq.~\ref{eq:rate2} and Eq.~\ref{eq:rate3} to be 7.8 kHz and 3.6 kHz respectively, representing a significant improvement over the current state of the art.

\begin{figure}
\includegraphics[width=1.0\columnwidth]{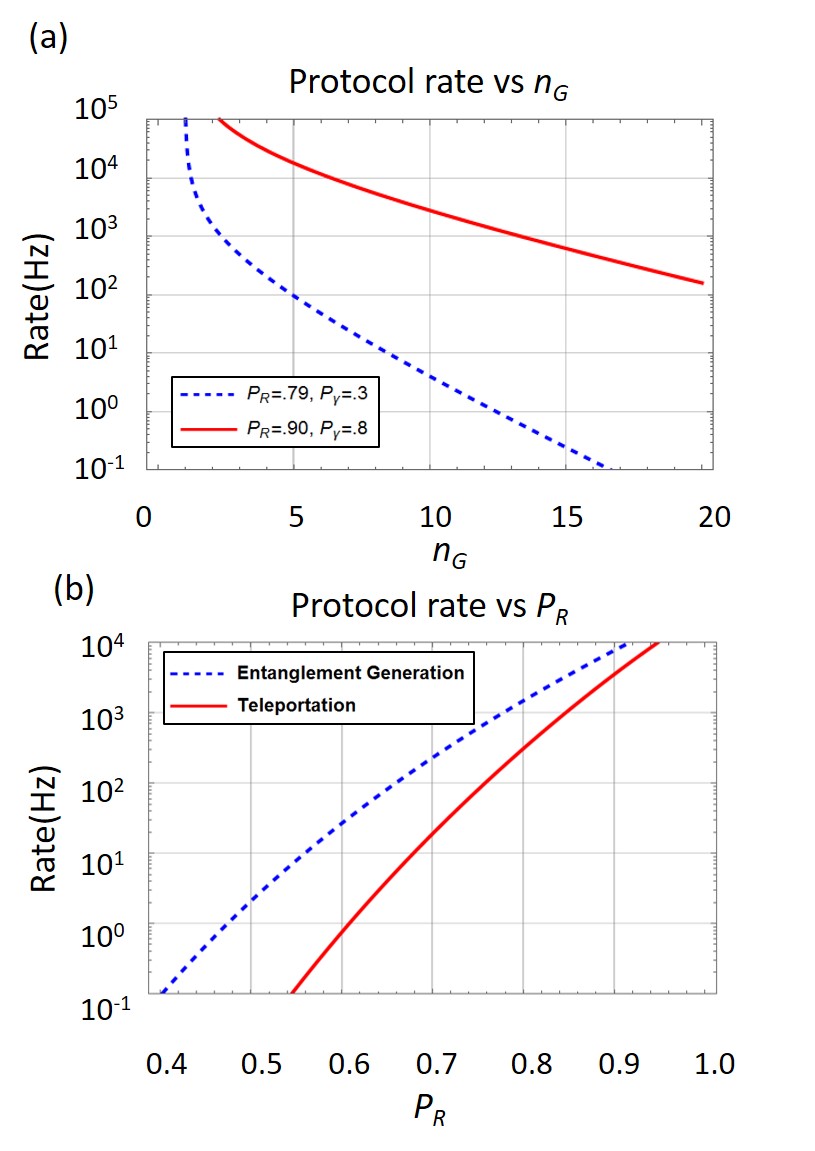}
\caption{\label{fig:rate} a) Plot of the rate of the entanglement generation protocol as a function of the number of steps in the protocol, $n_G$, for two sets of experimental parameters. The blue dashed line uses the conservativel efficiency estimates of $P_R = 0.79$ and $P_{\gamma}=0.3$ and the red solid line is the prediction using the optimistic efficiencies of $P_R = 0.9$ and $P_{\gamma}=0.8$. b) Plot of the rate of the entanglement generation protocol (dashed blue line) and teleportation protocol (solid red line) as functions of $P_R$ for a fixed $n_G = 7$ and $P_{\gamma}=0.8$.}
\end{figure} 

The rate of entanglement generation, i.e. Eq.~\ref{eq:rate3}, as a function of $n_G$ is plotted for the two different parameter sets mentioned in Fig.~\ref{fig:rate} (a). The rate of entanglement generation and teleportation for two-node protocols are compared using the optimistic photon collection efficiency $P_\gamma =0.8$, as a function of the Rydberg transition efficiency in Fig.~\ref{fig:rate} (b).

To compare to the rate of the Zhao {\sl et al.} protocol \cite{Zhao2010}, we note that for two nodes, the only difference is in the preparation of the flying qubit entangled with the quantum memory. In our case, $n_G =7$ whereas in Zhao {\sl et al.}, $n_G = 12$. As can be seen in Fig.~\ref{fig:rate}, this results in our protocol having a factor of $~\sim 20$ higher rate of entanglement generation than the Zhao {\sl et al.} protocol for the initial efficiency estimate $P_R = 0.79$, and a factor of 5 higher rate for the optimistic efficiency estimate $P_R = 0.9$. 

\section{\label{sec:level6}Many node rate analysis}
Next, we analyze the rates of these protocols in our model when extended to many equidistant nodes. Since the successful entanglement of two nodes is heralded by the detection of photons as described in Section~\ref{sec:level4}, the average time it takes to entangle two nodes can be used with a unity probability of success because we assume the entanglement of two nodes is successful every time the detection of the two photon state is heralded. The total average time to produce entanglement in the $n_{th}$ step, where $n$ is defined as one less than the number of nodes (because entanglement generation and teleportation are not defined for less than two nodes), is given by: 

\begin{eqnarray}
T_{G}[n]=\frac{T_{G}[n-1] + n_{S}t_o+ 2d/c}{(P_{R}^{n_{S}}P_{\gamma}^{2})^2P_{B}} \nonumber \\
=\frac{T_{G}[n-1]+ n_{S}t_o+ 2d/c}{P_{S}}
\label{eq:rate4}
\end{eqnarray}

Where $n_S$ is the number of additional atomic transitions required to prepare a subsequent flying qubit entangled with the quantum memory for the photonic entanglement swapping step described in Section~\ref{sec:level4}. In general $n_S \neq n_G$, as the number of steps to read out a memory that is already created is less than the number of steps required to produce a memory and read it out, but for the sake of simplicity, we will assume that $n_S \simeq n_G$ so that the total probability of successfully extending the entanglement is equal to the probability of generating entanglement between two nodes, i.e. $P_S=P_G$, which should be a good estimate for our purposes. 

If we set $P_R=1$ and use the average time to generate entanglement as the time for the first step, i.e. $T_{G}[1]=T_{G}$, we recover the logical solution of $T_G[n]=n T_G$. However, if instead we make the simplifying assumptions that the number of nodes is large and the probability of generating entanglement on a single shot is low, i.e. n $\gg 1$ and $P_G < 1$, then the solution is given by:

\begin{equation}
T_{G}[n]=\frac{t_G}{P_G^n(1-P_G)}
\label{eq:rate5}
\end{equation}

Because the entanglement swapping is not deterministic in this model, the protocols in Zhao {\sl et al.} \cite{Zhao2010} and Han {\sl et al.} \cite{Han2010}, which do have deterministic entanglement swapping, will outperform this one by a factor $ \mathcal{O}(P_G^n)$. 

The time for entanglement generation with deterministic entanglement swapping can be calculated to be \cite{Sangouard2009}:

\begin{equation}
T_{G}[n]=\Big(\frac{3}{2}\Big)^k\frac{t_G}{P_G}
\label{eq:rate5.5}
\end{equation}

Where k level of entanglement purification nesting such that $n = 2^k$, as before, $t_G$ is the time for the entanglement generation protocol, and $P_G$ is the probability of a successful entanglement event. Even for three nodes Eq.~\ref{eq:rate5.5} predicts an estimated improvement in the entanglement generation by a factor of 400 over Eq.~\ref{eq:rate5} for the initial efficiency estimates of $P_R = 0.79$ and $P_\gamma = 0.3$. For the improved efficiency estimates, $P_R = 0.90$ and $P_\gamma = 0.8$, the improvement using deterministic entanglement swapping is a factor of 100.

\begin{figure}
\includegraphics[width=1.0\columnwidth]{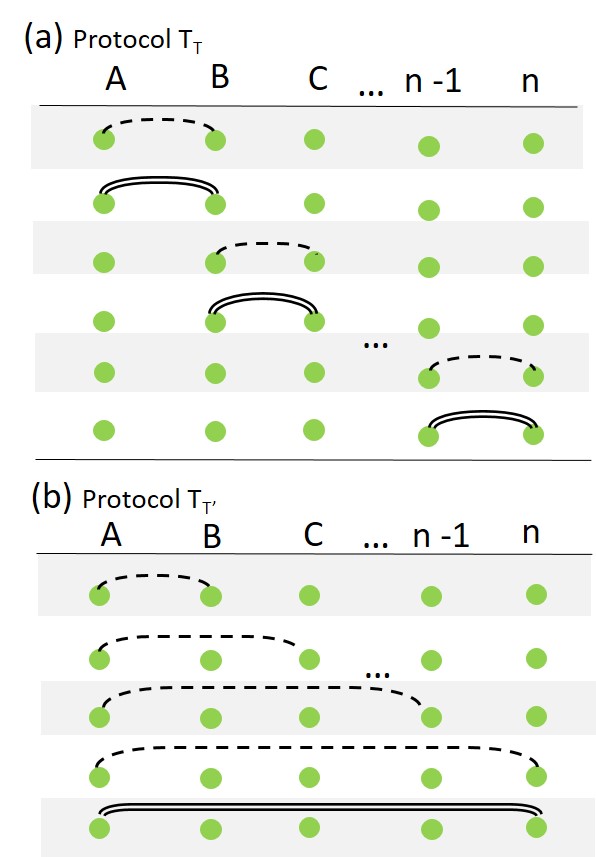}
\caption{\label{fig:tele}A dashed line represents establishing entanglement between two nodes and a double line represents teleporting a state between two nodes. a) A multi-node teleportation scheme (T) in which the entanglement is first generated between nodes and then the state is teleported. This followed by subsequently entangling nodes and teleporting the state until the final node is reached. b) A multi-node teleportation scheme (T') in which the entanglement is distributed from the first node to the last node by successive entangling operations. After the first and final node are entangled, the state is teleported.}
\end{figure} 
Next we analyze the rate of multi-node teleportation in the two cases shown in Fig.~\ref{fig:tele}. In the first case entanglement is generated between the first and second node. Then teleportation is performed on the target state to transfer it from the first to the second node. This is followed by subsequently entangling and teleporting the state down the chain until the target state is teleported to the final node, Fig.~\ref{fig:tele}(a). The rate can be calculated by assuming we have successfully teleported the state from the first node to the $n-1$ node. Then we solve for the intermediate step of generating the shared Bell-pair between the $n-1$ and $n_{th}$ nodes with a finite probability $P_G$:

\begin{equation}
\widetilde{T_{T}}[n]=\frac{T_{T}[n-1]+t_G}{P_G}
\label{eq:rate6}
\end{equation}

Once this is successfully completed, we perform the teleportation protocol given in Table~\ref{tab:table2}:

\begin{eqnarray}
T_{T}[n]=\frac{\widetilde{T_{T}}[n]+t_T}{P_T} \nonumber \\
=\frac{T_{T}[n-1]+t_G}{P_TP_G}+\frac{t_T}{P_T} 
\label{eq:rate7}
\end{eqnarray}

To solve this, we make the simplifying assumptions that the number of steps is very large, n $\gg 1$ and $P_T,P_G < 1$ so that $(P_TP_G)^n \ll 1$ and that the time it takes for teleportation from the first to second node is given by $T_T$ from Eq.~\ref{eq:rate3}. The solution for large $n$ simplifies to:
\begin{equation}
T_{T}[n]=\frac{t_G}{(P_GP_T)^{n}(1-P_GP_T)}
\label{eq:rate8}
\end{equation}

Next, we want to analyze the rate of the teleportation scheme shown in Fig.~\ref{fig:tele}(b). For this scheme, the entanglement is generated from one node to the last node, followed by a single teleportation step. The time it takes to do this is given by Eq.~\ref{eq:rate5}. This is followed by a single teleportation step. If we take $T_G[n]$ as the first step, and assume that the time for the teleportation protocol is negligible compared to the time of the entanglement generation of all of the nodes, we simply have $T_{T'}[n]=T_G[n]/P_T$ or:
\begin{equation}
T_{T'}[n]=\frac{t_G}{P_G^{n}P_T(1-P_G)}
\label{eq:rate9}
\end{equation}
This is significantly faster than the rate of Protocol $T_T$, i.e. Eq.~\ref{eq:rate8}, by a factor of $(P_R^{n_T})^n$, and could in fact be improved with a nested entanglement quantum repeater protocol, which is not possible with protocol $T_T$. 

If we use the optimistic efficiency estimates, i.e. $P_R = 0.9$ and $P_\gamma = 0.8$, then Eq.~\ref{eq:rate8} and Eq.~\ref{eq:rate9} predict an average time to teleport a state between three nodes, i.e. from node A to node C, of 145 ms for protocol $T_T$ and 34 ms for protocol $T_{T'}$. Protocol $T_{T'}$ predicts an average time of 90 s to teleport a state to the 6th node where protocol $T_{T}$ predicts an average time of around one hour.

However, protocol $T_{T'}$, Fig.~\ref{fig:tele}(b) requires that the memory lifetime of the first node to be long enough for the entire protocol to be successful, whereas protocol $T_T$, Fig.~\ref{fig:tele}(a) only requires a memory time long enough to teleport a state between neighboring nodes. In addition, protocol $T_{T}$ is more resource intensive, requiring the `target' qubit pair used in the teleportation protocol at each node, while the $ T_{T'}$ protocol only requires the `target' pair at the initial node. 

\section*{CONCLUSIONS}
Using a multi-mode Rydberg excitation scheme in an atomic ensemble, teleportation between long-lived memory states can achieve high rates. This system has also been shown to be compatible with a large-scale quantum network architecture. We have examined the performance of a quantum repeater node based on cold-atom ensembles with Rydberg excitations and theoretically described a teleportation protocol. We analyzed the rates of two-node entanglement generation and teleportation and found that the teleportation rates achievable on realistic systems could approach the kHz level, two orders of magnitude improvement over the current highest achieved rate. This two-node performance can be used as a metric and benchmark for incorporating Rydberg-based cold-atom ensemble quantum repeater nodes into a larger scale network. We also analyzed a model for many node protocols.

It could be possible to spatially multiplex a cold atom ensemble node by addressing several Rydberg radii of atoms along the length of the optical dipole trap or by multi-site trapping of an atomic ensemble on a chip with individual site addressing technology \cite{Lan2009}. The technology for coherent control of Rydberg atoms on a chip, though challenging, is currently being pursued by several groups \cite{Leung2014,Hermann-Avigliano2014} where it might be possible to realize multi-Rydberg atom trapping on atom chips. These types of multi-plexing could realize larger quantum registers at each node and would enable temporal multiplexing to increase data transmission rates. 

The presence of a single Rydberg excitation produces a large non-linearity in the ensemble which can affect other atoms in the ensemble or photons entering the ensemble. This can lead to non-linear effects at the single photon level such as single photon EIT\cite{Pritchard2010}, single photon switches\cite{Peyronel2012, Baur2014}, single photon transistors\cite{Gorniaczyk2014}, and effective photon-photon interactions\cite{Firstenberg2013} which have all been experimentally demonstrated. Rydberg excitations in cold atom ensembles promise a rich and viable path towards interesting applications in quantum communication and information.

Because of the potential high rates of entanglement generation and the potential for scalability, quantum repeaters based on neutral atom ensembles with Rydberg excitations are a promising route towards long-distance quantum network.

\section*{Acknowledgments}
We would like to thank Steve Rolston and Alexey Gorshkov for helpful discussions. NS is an Oak Ridge Associated Universities (ORAU) postdoctoral fellow.
Research was sponsored by the Army Research Laboratory. The views and
conclusions contained in this document are those of the Authors and should
not be interpreted as representing the official policies, either expressed
or implied, of the Army Research Laboratory or the U.S. Government. The U.S.
Government is authorized to reproduce and distribute reprints for Government
purposes notwithstanding any copyright notation herein.

\end{document}